\documentclass{article}
\usepackage{graphicx}
\usepackage{epsfig}
\usepackage{url}
\usepackage{amssymb}
\usepackage{amsmath}
\usepackage{bm}
\usepackage[numbers]{natbib}
\usepackage{10thHSTAM}

\newcommand{\vect}[1]{\boldsymbol{#1}}

\begin{document}
\title{HIGH ORDER THREE PART SPLIT SYMPLECTIC INTEGRATION SCHEMES}
\author{Enrico Gerlach$^1$, Siegfried Eggl$^2$, Charalampos Skokos$^3$, Joshua~D. Bodyfelt$^4$ and Georgios Papamikos$^5$}
\maketitle
\address{$^1$Lohrmann Observatory, Technical University Dresden, D-01062, Dresden, Germany, enrico.gerlach@tu-dresden.de}
\address{$^2$IMCCE, Observatoire de Paris, 77 Avenue Denfert-Rochereau , F-75014, Paris, France, siegfried.eggl@imcce.fr}
\address{$^3$Physics Department, Aristotle University of Thessaloniki, GR-54124, Thessaloniki, Greece, hskokos@auth.gr}
\address{$^4$The Ohio State University, ElectroScience Laboratory, 1320 Kinnear Road, Columbus, OH 43212, USA, jdbodyfelt@gmail.com}
\address{$^5$School of Mathematics, Statistics and Actuarial Science, University of Kent, Canterbury, CT2 7NF, UK, geopap1983@gmail.com}


\begin{abstract}
Symplectic integration methods based on operator splitting are well established in many branches of science.
For Hamiltonian systems which split in more than two parts, symplectic methods of higher order have been studied in detail only for a few special cases.
In this work, we present and compare different ways to construct high order symplectic schemes for general Hamiltonian systems
that can be split in three integrable parts. We use these techniques to numerically solve the equations of motion for a simple toy model, as well as
the disordered discrete nonlinear Schr\"odinger equation. We thereby compare the efficiency of symplectic and non-symplectic integration methods.
Our results show that the new symplectic schemes are superior to the other tested methods, with respect to both long term energy conservation and
computational time requirements.
\end{abstract}

\section{INTRODUCTION}
When studying mechanical systems, a Hamiltonian formulation of problems is often advantageous, since equations of motion can be easily obtained
from a scalar function $H$ representing the total energy of the system. The resulting set of first-order differential equations describes the time evolution of the dependent variables on a differentiable manifold. In what follows next we deal with the case where the phase space is an Euclidean space $\mathbb R^{2N}$ with the coordinates being the generalized positions $q_l$ and momenta $p_l$ with  $l=1,2,\ldots,N$ and $N$ being the number of the system's degrees of freedom. Except for a few special cases, where the solution of the equations of motion can be written in a closed analytic form,
the system's trajectories in phase space must be approximated by numerical means.

Due to their excellent performance, especially over long integration times, the so-called symplectic integration techniques are of particular interest for the numerical integration of Hamiltonian systems. Numerical integration algorithms showing symplectic properties can be traced back to Isaac Newton's `Principia Mathematica' (1687) \citep{hairer-et-al-2006}. The main reason why symplectic integrators (SIs) have become so popular over the past decades lies in the fact that they offer remarkable long-term energy conservation. In fact, the local error in the system's total energy does not grow with time as is the case for most non-symplectic methods \citep[see e.g.][]{eggl-dvorak-2010}. The underlying reasons for the excellent performance of symplectic integrators in this respect were understood only in the late $20^{th}$ century \citep[see e.g.][and references therein]{yoshida-1993,hairer-et-al-2006}. Recently, it was shown that SIs are also highly efficient in the integration of the variational equations
needed for the computation of chaos indicators like the maximum Lyapunov Characteristic Exponent (mLCE) \citep[see e.g.][]{S2010}, the Smaller (SALI) \citep{S2001, SABV2003, SABV2004} and Generalized Alignment Index (GALI) \citep{SBA2007} when using the so-called `Tangent Map' method \citep{SG2010, GS2011, GES2012}. Due to these benefits, SIs have become a standard technique in Hamiltonian dynamics with particular importance in long-term integrations of multidimensional systems.

In cases where the Hamiltonian $H$ can be separated in two parts $H = H_A + H_B$, which offer individually integrable state transition maps, various symplectic integrators have been developed over the past years. For an overview see \citep{hairer-et-al-2006} and references therein.
However, in many physical problems the Hamiltonian cannot be split in merely two separable parts. In this paper we focus on general Hamiltonian systems that can be split in exactly three integrable parts.
We will show how high order symplectic integration methods for these kinds of systems can be constructed.
Using two different Hamiltonians as examples, we will compare triple split methods with respect to their computational efficiency and energy conservation.

This paper is structured as follows: in section \ref{sec:SI} we give a brief introduction to the theory of SIs. In section \ref{sec:3part} we present a systematic way to construct high order SIs for three part Hamiltonian systems.
We  apply these methods and compare their efficiency using a toy model (section \ref{sec:toy}) as well as the disordered discrete nonlinear Schr\"odinger equation (section \ref{sec:DNLS}) as showcases. In section \ref{sec:conclusion} we summarize our results.

\section{\label{sec:SI}SYMPLECTIC INTEGRATION OF HAMILTONIAN SYSTEMS}
Finding and solving the equations of motion for dynamical systems can be greatly simplified,
when a description in a Hamiltonian framework is possible. Given a twice continuously differentiable function $H(\vect{z})$ which represents
the total mechanical energy of a system, the equations of motion simply derive from
\begin{equation}
\frac{d}{dt}\vect{z}=\vect{J} \nabla_{\vect{z}} H,  \qquad \text{where}
\qquad \vect{J}=\left(
\begin{array}{cc}
 \bm 0 & \vect I\\
  -\vect I &\bm 0\\
\end{array} \right), \label{eq:ham}
\end{equation}
and $\vect z = (\vect{q},\vect{p})^T$ represents the vector of generalized coordinates and momenta. Furthermore, $\vect I$ is the $N$-dimensional identity matrix, and $\bm 0$ the $N \times N$ matrix with all its elements equal to zero, with
$N$ being the number of degrees of freedom of the dynamical system.
Differential equations derived from Hamiltonian systems possess a special geometric quality related to the matrix $\vect{J}$ - they are `symplectic'.
To be more precise, the continuous flow of the system, i.e.~the continuous function $\phi_t:\vect{z}(0)\rightarrow\vect{z}(t)$, which maps initial conditions into system states at time $t$,
keeps the symplectic geometric structure $\vect J$ intact \citep[for a proof see e.g.][]{hairer-et-al-2006}.
In analogy, discrete integration algorithms that conserve $\vect J$ are called `symplectic' as well.

Let us consider the example of the so-called `symplectic Euler' integration method applied to the well known harmonic oscillator problem. Here, $H(q,p)=(q^2+p^2)/2$, $N=1$ and the flow map $\phi_t$, which is the action of the Lie group $SO(2)$ on the plane $\mathbb{R}^2$, is approximated by the symplectic Euler scheme
\begin{eqnarray}
\label{eq:syeul}
q_\tau &=& q_0+\tau \partial_p H|_{q_0,p_0} = q_0+\tau p_0 \nonumber \\
p_\tau &=& p_0-\tau \partial_q H|_{q_\tau,p_0} =p_0-\tau q_\tau,
\end{eqnarray}
where $\tau$ is the integration time step. The numerical integration method is explicit and it is defined by the state transition matrix
$\vect \Phi$
\begin{equation}
\vect\Phi=\left(\begin{array}{cc}      1& \tau\\
-\tau& 1-\tau^2\\  \end{array}\right).
\end{equation}
A straightforward calculation shows that
\begin{equation}
\vect \Phi\; \vect J\; \vect \Phi^T= \vect J,
\end{equation}
which means that $\vect \Phi$ preserves $\vect J$. Therefore, the integration method presented in equations (\ref{eq:syeul}) is indeed symplectic.

There are many possible ways of constructing symplectic integrators, but operator splitting is one of the
most transparent concepts.
We slightly reformulate Hamilton's equations of motion (\ref{eq:ham})
\begin{equation}
\frac{d}{dt}\vect{z}=\vect{J} \nabla_{\vect{z}} H = \{H,\vect z\} = L_H \vect z \label{eq:lie}
\end{equation}
where $L_H=\{H,\cdot\}$ is a differential operator and  $\{\cdot,\cdot\}$ is the Poisson bracket\footnote{In this paper we always refer to the canonical Poisson bracket
$\{f,g\}=\sum_{i=1}^N \frac{\partial f}{\partial p_i}\frac{\partial g}{\partial q_i} - \frac{\partial f}{\partial q_i}\frac{\partial g}{\partial p_i}$.
}.
Note that the symplectic structure is now contained implicitly in the Poisson bracket.
The formal solution of differential equations (\ref{eq:lie}) reads
\begin{equation}
\vect{z}(\tau)=e^{\tau L_H}\vect z (0) \label{eq:exp}.
\end{equation}
If the Hamiltonian can be written as a sum of functions, e.g. $H=\sum_{i=1}^m H_i$,
the bilinearity of the Poisson bracket allows us to rewrite equation (\ref{eq:exp})
\begin{equation}
\vect{z}(\tau)=e^{\tau L_H}\vect z (0)= e^{\tau \sum_i^m L_{H_i}}\vect z(0).
\end{equation}
The combined map $\phi_\tau:\vect z (0) \rightarrow \vect z (\tau)$, $\phi_\tau=e^{\tau \sum_i^m L_{H_i}}$ usually does not permit an analytic solution, but the individual exponential maps $\phi_\tau^i=e^{\tau L_{H_i}}$ might.
In the case of $H(\vect z)=H_A(\vect p)+H_B(\vect q)$ the operator splitting is then executed as follows
\begin{equation}
\vect{z}(\tau)= e^{\tau L_H}\vect z(0)=e^{\tau(L_A+L_B)}\vect z(0)\approx e^{\tau L_A}e^{\tau L_B}\vect z(0). \label{eq:split}
\end{equation}
In our example, analytic solutions can indeed be found for the individual maps, since $H_A$ depends only on the generalized momenta $\vect p$, representing e.g.~the total kinetic energy of the system. Similarly, $H_B$ is a pure function of the generalized coordinates, e.g.~the potential energy.
Such an approach is suitable for creating explicit integration algorithms of first order in $\tau$. For higher order integrators that permit larger time steps
we will have to take a more careful look at the approximation in equation (\ref{eq:split}). In fact, one can show that the operator splitting
produces error terms of higher order in $\tau$ due to the Baker-Campbell-Hausdorff (BCH) relation \citep{hairer-et-al-2006}
\begin{equation}
\begin{tiny}
e^{\tau L_A}e^{\tau L_B} = e^{\tau(L_A+L_B)+\frac{\tau^2}{2}[L_A,L_B]+\frac{\tau^3}{12}([L_A,[L_A,L_B]]-[L_B,[L_A,L_B]])+\hdots},\label{eq:bch}
\end{tiny}
\end{equation}
where $[\cdot,\cdot]$ denote commutators for the operators $L_A$ and $L_B$, i.e.~$[L_A,L_B]=L_AL_B-L_BL_A$. This representation has lead \citep{yoshida-1993} to the conclusion that, in fact, not the intended system, but a dynamical system close to the original one is solved exactly by the split maps. For example it is easy to see that for the harmonic oscillator, the corresponding symplectic Euler map
\begin{equation}
\left(
\begin{array}[h]{c}
	q_{\tau} \\
	p_{\tau}
\end{array}
\right)=\left(
\begin{array}[h]{cc}
	1 & \tau \\
	-\tau & 1-\tau^2
\end{array}
\right)\left(
\begin{array}[h]{c}
	q_0\\
	p_0
\end{array}
\right)
\end{equation}
leaves the ellipse $I_{\tau}=(q^2+p^2+\tau qp)/2=H+\tau qp/2$ invariant and thus it preserves exactly a perturbed energy.
As can be seen from equation (\ref{eq:bch}), the difference between the original and the nearby system is a polynomial in $\tau$ with the commutators of the split operators as coefficients. While the exact solution of a nearby dynamical system explains the bounded local error in energy, equation (\ref{eq:bch}) also contains the reason why variable time stepping destroys the favorable energy conservation properties of symplectic algorithms.
Changing the time step means changing the analytically solved nearby Hamiltonian continuously. Consequently, the advantage of solving a
system which stays in the vicinity of the original one is lost.

The standard procedure of constructing methods of higher order $p$ usually consists of trying to find numerical coefficients $a_i$ and $b_i$,
so that a sequence of split maps eliminates consecutive terms of the BCH expansion
\begin{equation}
e^{\tau(L_A+L_B)}=\prod_{i=1}^p e^{\tau a_i L_A}e^{\tau b_i L_B}+O(\tau^{p+1}),
\end{equation}
see for instance \citep{candy-rozmus-1991,suzuki-1992,gray-manolopoulos-1996}. The well-known leap-frog (aka St\"ormer/Verlet) method of order two is obtained in this way by setting $a_1=1/2$ and $b_1=1$. It can be shown \citep{suzuki-1991} that it is not possible to construct integrators of order $p>2$ having only positive steps.
Since negative steps limit the stability of the algorithm, it is tempting to circumvent this problem.
For the special case of nearly integrable systems of the form $H=H_A + \epsilon H_B$ with $\epsilon\ll 1$
very efficient symplectic integrators with only positive steps can be constructed, as is shown in \citep{LR2001}.

\section{\label{sec:3part}SYMPLECTIC THREE PART SPLITTING SCHEMES}

Let us now consider general Hamiltonians that can be split in three parts $H=H_A+H_B+H_C$, where each of the maps $e^{\tau L_A}$, $e^{\tau L_B}$ and $e^{\tau L_C}$ has an analytic solution. Then the construction of a symplectic integration method can be achieved as follows
\begin{equation}
e^{\tau(L_A+L_B+L_C)}\approx \prod_{i=1}^p e^{\tau a_i L_A}e^{\tau b_i L_B}e^{\tau c_i L_C}.
\end{equation}
The simplest integration scheme is of first order and is just the concatenation $e^{\tau(L_A+L_B+L_C)}\approx e^{\tau L_A}e^{\tau L_B}e^{\tau L_C}$. In composition with its adjoint method (which is also symplectic) this yields already a time-reversible scheme of order two, which we will call $ABC_2$, i.e.
\begin{equation}
\label{eq:ABC2}
 ABC_2 := e^{\tau/2 L_A}e^{\tau/2 L_B}e^{\tau L_C}e^{\tau/2 L_B}e^{\tau/2 L_A}.
\end{equation}
This integrator has already been applied for the numerical study of astronomical problems \cite{C_99,GBB_08,C_10}. Some sporadic attempts to construct higher order three part split SIs for specific dynamical systems have also been performed.
Advanced methods, for instance, especially designed for molecular dynamics are given in \citep{omelyan-2007}, while optimized algorithms for highly accurate long-term integration of astronomical problems are presented in \citep{farres-2012}. Recently, an attempt to systematically construct high order three part SIs was carried out in \cite{DNLS}. In that paper, several SIs were presented for the integration of a Hamiltonian system that splits in three integrable parts. In addition, the performance of these numerical schemes for the integration of a multidimensional Hamiltonian system of particular physical interest was studied.
In the following subsections we will discuss various approaches to integrate a general three part Hamiltonian problem following the ideas presented in \cite{DNLS}.

\subsection{Yoshida composition method}
In \citep{yoshida-1990} it was shown that the concatenation of time symmetric splittings can produce higher order methods using analytically derived coefficients only. With this approach one can construct a SI: $\mathcal S_{p+2}(\tau)$ of order $p+2$ starting from a SI: $\mathcal S_p(\tau)$ of even order $p$ by
\begin{equation}
\label{eq:yoshida}
 \mathcal S_{p+2}(\tau) = \mathcal S_p(x_1\tau)\mathcal S_p(x_0\tau)\mathcal S_p(x_1\tau),
\end{equation}
with
\begin{equation}
\label{eq:yoshida_coef}
x_0=-\frac{2^{1/(2p+1)}}{2-2^{1/(2p+1)}} , \,\,\,\, x_1=-\frac{1}{2-2^{1/(2p+1)}}.
\end{equation}
Applying this procedure to the second order $ABC_2$ method (\ref{eq:ABC2}) for $p=2$, one obtains a SI of order 4 with 13 steps, which will be called $ABC_4$
\begin{equation}
 ABC_4 := ABC_2(x_1\tau) ABC_2(x_0\tau) ABC_2 (x_1\tau).
\end{equation}
Starting from $ABC_4$ and using equation (\ref{eq:yoshida}) again, one could continue and build a method of order 6. Although this procedure is straightforward, it is not optimal with respect to the number of required steps, i.e. force evaluations per time step. As was already pointed out in \citep{yoshida-1990}, alternative methods can be applied to obtain more economical integrators of high order, although the new coefficients cannot be given in
analytical form. In this work we use a method of order 6 having in total 29 steps
with
\begin{equation}
 ABC_6 := ABC_2(w_3\tau) ABC_2(w_2\tau) ABC_2 (w_1\tau)ABC_2(w_0\tau) ABC_2(w_1\tau) ABC_2 (w_2\tau)ABC_2(w_3\tau).
\end{equation}
The exact values of $w_i$ for $i=0,1,2,3$ can be found in \citep[Chapter V, equation (3.11)]{hairer-et-al-2006} and \citep{yoshida-1990}.

\subsection{Successive splittings}
A second possibility to obtain SIs for problems where $H=H_A+H_B+H_C$ is to split the Hamiltonian in two parts as $\mathcal A=A$ and $\mathcal B=B+C$ and apply one of the various available symplectic two part algorithms
\begin{equation}
e^{\tau(L_{\mathcal A}+L_{\mathcal B})}\approx \prod_{i=1}^p e^{\tau a_i L_A}e^{\tau b_i L_{B+C}}.
\end{equation}
In a successive splitting $e^{\tau b_i L_{B+C}}$ is approximated by an appropriate numerical scheme.
To give an example: the application of the traditional leap-frog method will lead in a first step to $e^{\tau/2 L_\mathcal A} e^{\tau L_\mathcal B} e^{\tau/2 L_\mathcal A}$.
In a further splitting $e^{\tau L_\mathcal B}$ is approximated again by a leap-frog step to give a second order symplectic integrator $\mathcal S_2$ with
\begin{equation}
 \mathcal S_2(\tau) = e^{\tau/2 L_A} \left[e^{\tau/2 L_B} e^{\tau L_C} e^{\tau/2 L_B} \right]e^{\tau/2 L_A},
\end{equation}
where the brackets $[\cdot]$ here are used only to emphasize the structure.
We note that $\mathcal S_2$ obtained by successive splitting of leap-frog steps is identical to $ABC_2$ (see equation (\ref{eq:ABC2})).
Using different splitting techniques will generally lead to methods with various strengths and weaknesses. In \citep{LR2001}, for instance,
an integrator of second order called $SABA_2$ was introduced,
which showed improved efficiency compared to the traditional leap-frog at only slightly larger computational cost. Applying a similar splitting to our problem we obtain the integrator $e^{a_1\tau L_\mathcal A} e^{b_1\tau L_\mathcal B}e^{a_2\tau L_\mathcal A}e^{b_1\tau L_\mathcal B}e^{a_1 L_\mathcal A}$, with $a_1=(3-\sqrt{3})/6, a_2=\sqrt{3}/3$ and $b_1=1/2$. The approximation of $e^{b_1\tau L_\mathcal B}$ again with the $SABA_2$ scheme leads to an integrator with 13 steps, which we call $SS_2$.

Each of these second order methods obtained by successive splitting can be combined into an integrator of fourth order by application of equation (\ref{eq:yoshida}).
Thus, one can construct $SS_4$ having 37 simple steps from concatenations of $SS_2$.
In general, the number of steps for $SS_p$ grows quickly, which makes successive splitting methods computationally expensive compared to $ABC_p$ integrators of the same order.

\subsection{Non-symplectic schemes}
Of course, the numerical solution of the equations of motion for a given Hamiltonian problem does not necessarily have to be performed with symplectic schemes.
In principle, one can use any general-purpose non-symplectic integration scheme for this task.
The disadvantage of using non-symplectic algorithms is that the error in the total energy of the system grows with time. Therefore, different epochs of the system's evolution are computed with different accuracy. Especially for problems where the asymptotic behavior of the system in the later stages of the evolution is of main interest, this represents a serious drawback.
Nevertheless, we will compare the performance of symplectic algorithms to solutions obtained using the `classic' explicit Runge-Kutta method of order 4 ($RK_4$) with a fixed time step.

\subsection{\label{sec:toy}A toy model}
As a first step in testing the efficiency of the different integration methods we apply our splitting procedures to a simple toy model described by the following Hamiltonian
\begin{equation}
 H = \frac{1}{2}\left(p^2 + q^2 + p^2q^2 \right),
\label{eq:testH}
\end{equation}
where we set $H_A=p^2/2,\ H_B=q^2/2$ and $H_C=p^2q^2/2$. For this ordering one can identify the map $\phi_\tau$ over one time step $\tau$ from $(q,p)$ to $(q',p')$ as
\begin{equation}
e^{\tau L_A}: \left\{
\begin{array}{lll}
  q' & = & q + \tau p\\
  p' & = & p \\
\end{array}\right. ,
\end{equation}
\begin{equation}
e^{\tau L_B}: \left\{
\begin{array}{lll}
q' & = & q \\
p' & = & p - \tau q\\
\end{array}\right. ,
\end{equation}
\begin{equation}
e^{\tau L_C}: \left\{
\begin{array}{lll}
q' &=& q e^{\tau pq}\\
p' &=& p e^{-\tau pq}\\
\end{array}\right. .
\end{equation}

\begin{figure}[h]	
\centerline{
\includegraphics*[width=0.6\textwidth]{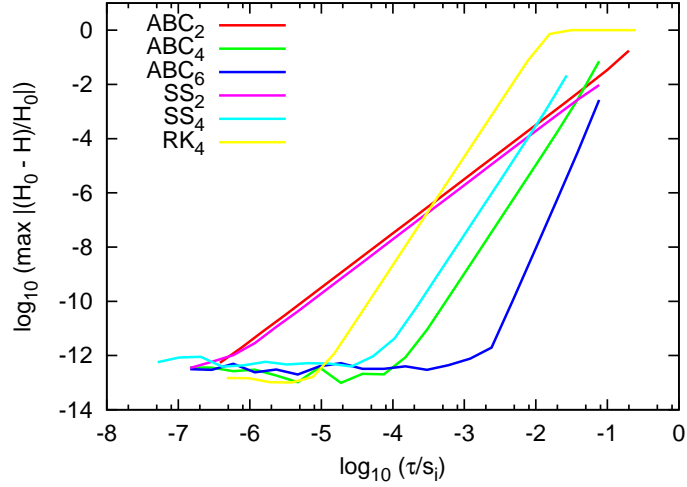}}
\caption{Comparison of the results obtained with a non-symplectic ($RK_4$) and different symplectic three part split schemes for the integration of the Hamiltonian (\ref{eq:testH}). The horizontal axis gives the computational gain $\tau/s_i$, while the maximum value of the absolute relative energy variation over $10^6$ time steps is shown on the vertical axis.}
\label{fig:1}
\end{figure}

In our tests we will compare the various methods presented earlier with regards to efficiency and energy conservation.
In order to complete one time step $\tau$ each method requires a different number of evaluations of $e^{\tau L_{A,B,C}}$.
The total number of these evaluations gives the number of steps $s_i$ characteristic for each method.
It is clear that $s_i$ will be larger for higher order methods, since more evaluations per time step are generally required.
On the other hand, the advantage of using higher order methods lies in that one usually gets better precision for a given time step.
If a method $\mathcal S_a$ needs twice the number of evaluations compared to another method $\mathcal S_b$, but the increase in precision is larger than a factor of two, we say
that $\mathcal S_a$ is more efficient than $\mathcal S_b$.
Accordingly, our tests will measure the computational gain of a method $\mathcal S_i$ by $gain=\tau/s_i$, where $s_i$ is the number of evaluations the respective scheme needs for one time step $\tau$.
This computational gain will be compared to the precision that can be achieved with a given time step $\tau$. In order to measure the precision of a specific integration method, we use the maximum value of the absolute relative error in energy $\max |(H(t_0)-H(t))/H(t_0)|$ accumulated over $10^6$ time steps.
A performance comparison of the different methods presented in the previous sections is given in Figure \ref{fig:1}.

As expected, methods of the same order show a similar behavior, i.e.~the same slope of increase in precision when the time step $\tau$ is decreased. We see that both methods of order two give almost identical results. Interestingly, $ABC_2$ (red line) performs slightly worse than $SS_2$ (pink line) despite the considerably lower number of force evaluations per step ($s_{SS_2}=13$ and $s_{ABC_2}=5$).
For the fourth order methods, however, we see this trend reversed. With $ABC_4$ (green line) one obtains the best performance of all fourth order schemes.
Furthermore, both symplectic algorithms provide stable results for relatively large time steps, which is not the case for $RK_4$ (yellow line).
The best overall performance is achieved by using the method with the highest order: $ABC_6$ (blue line).
Although this method needs 29 force evaluations during each time step, it outranks other schemes in terms of computational gain for a given precision and quickly reaches the round-off limit in energy conservation.

\subsection{\label{sec:DNLS}The disordered discrete nonlinear Schr\"odinger equation}
In order to investigate the efficiency of the different SIs in a more elaborate setting, we follow \cite{DNLS} and consider a one-dimensional chain of coupled, nonlinear oscillators: the disordered discrete nonlinear Schr\"odinger equation (DNLS).
This system is   described by the Hamiltonian
\begin{equation}
\mathcal{H}_{D}= \sum_{l} \epsilon_{l}
|\psi_{l}|^2+\frac{\beta}{2} |\psi_{l}|^{4}
- (\psi_{l+1}\psi_l^*  +\psi_{l+1}^* \psi_l),
\label{RDNLS}
\end{equation}
with complex variables $\psi_{l}$, lattice site indices $l$ and
nonlinearity strength $\beta \geq 0$.  The random on--site energies
$\epsilon_{l}$ are chosen uniformly from the interval
$\left[-W/2,W/2\right]$, where $W$ denotes the
disorder strength. This model has two integrals of motion as it
conserves the energy (\ref{RDNLS}) and the norm $S =
\sum_{l}|\psi_l|^2$, and  has been extensively studied to determine the characteristics of energy spreading in disordered systems \citep{KKFA2008,FKS09,SKKF09,LBKSF10,BLGKSF2011,BLSKF11}. It was shown in these studies that the second moment of the norm distribution, $m_2$, grows subdiffusively in time $t$, namely as $t^a$. The asymptotic value $a=1/3$ of the exponent was theoretically predicted and numerically
verified.

The final fate of wave packets in multidimensional disordered lattices is a highly debatable physical problem, however. What will happen when $t \rightarrow \infty$?
Will wave packets continue spreading for ever, as  numerical simulations suggest, or will the spreading eventually stop, as claimed by some researchers \cite{JKA_10,A_11}?
In order to numerically tackle these questions, we need efficient integration schemes,
which allow accurate integrations for times longer than the ones achieved by currently existing means.
In what follows, we will have a closer look at the promising symplectic methods that were introduced in \cite{DNLS}.

The canonical transformation
\mbox{$\psi_l=(q_l+ip_l)/\sqrt{2}$},
\mbox{$\psi_l^*=(q_l-ip_l)/\sqrt{2}$}, gives Hamiltonian (\ref{RDNLS}) the form
\begin{equation}
  H_D=\sum_l \frac{\epsilon_l}{2}(q_l^2+p_l^2)+\dfrac{\beta}{8}(q_l^2+p_l^2)^2 - p_{l+1} p_l -q_{l+1}q_l,
\label{HDNLS}
\end{equation}
where $q_l$ and $p_l$ are generalized coordinates and
momenta, respectively. We split Hamiltonian (\ref{HDNLS})
into a sum of three parts, namely  $H_A=\sum_l \frac{\epsilon_l}{2}(q_l^2+p_l^2)+\dfrac{\beta}{8}(q_l^2+p_l^2)^2$, $H_B=-\sum_l  p_{l+1} p_l$ and $H_C=-\sum_l q_{l+1}q_l$.
The corresponding exponential maps over one time step $\tau$ are
\begin{equation}
e^{\tau L_A}: \left\{ \begin{array}{lll} q'_l & = & q_l \cos(\alpha_l \tau)+ p_l
\sin(\alpha_l \tau)\\ p'_l & =& p_l \cos(\alpha_l \tau)- q_l \sin(\alpha_l
\tau) \\
\end{array}\right. ,
\label{eq:LA}
\end{equation}
\begin{equation}
e^{\tau L_B}: \left\{ \begin{array}{lll} p'_l & =& p_l \\ q'_l & = &
q_l-(p_{l-1}+p_{l+1}) \tau \\
\end{array}\right. ,
\label{eq:LB}
\end{equation}
\begin{equation}
e^{\tau L_C}: \left\{ \begin{array}{lll} q'_l & =& q_l \\ p'_l & = &
p_l+(q_{l-1}+q_{l+1}) \tau \\
\end{array}\right. ,
\label{eq:LC}
\end{equation}
with $\alpha_l=\epsilon_l+\beta(q_l^2+p_l^2)/2$. Thus, the DNLS model
represents an ideal `real-world' test case for the aforementioned three part split
SIs.

\begin{figure}[h]	
\centerline{
\includegraphics*[width=0.7\textwidth]{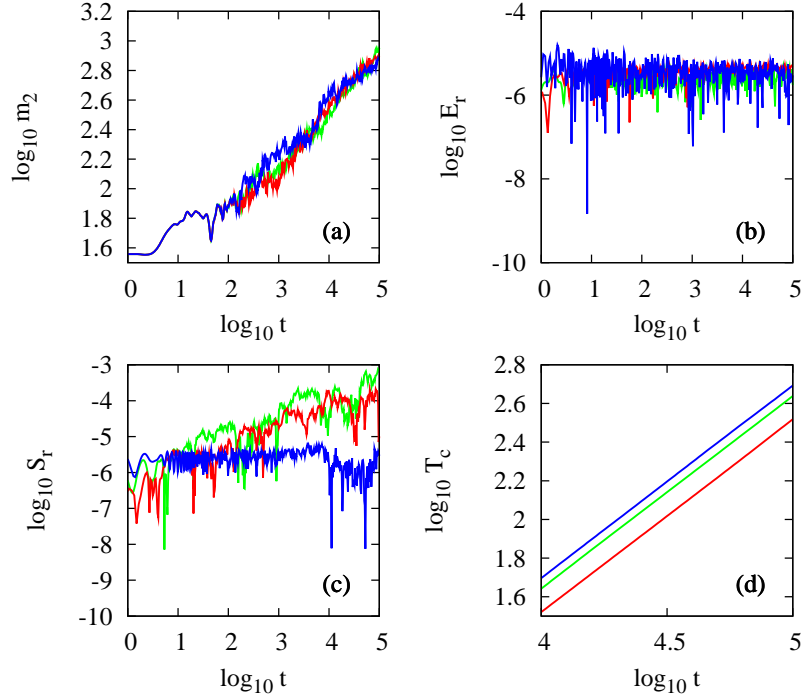}}
\caption{Results for the integration of $H_D$ (\ref{HDNLS}), by the fourth order SIs $SS_4$ for $\tau=0.1$ (blue line),
$ABC_4$ with $\tau=0.05$ (green line) and the sixth order $ABC_6$ for $\tau=0.15$ (red line). The different panels show the time evolution of
(a) the logarithm of the second moment $m_2(t)$, (b) the absolute relative energy error $E_r(t)$,
(c) the absolute relative norm error $S_r(t)$ and (d) the required CPU time $T_c(t)$ in seconds. See text for details.}
\label{fig:2}
\end{figure}

To compare the performance of the various integration schemes
we consider a particular disorder realization of the DNLS model
(\ref{HDNLS}) with $1024$ lattice sites $(l=1,\hdots, 1024)$ and fixed boundary conditions. We fix the total norm of
the system to $S=1$, and, following \cite{LBKSF10}, we excite
 21 central sites homogeneously by attributing the
same constant norm to every individual site with a random initial phase. For all other sites
we set $q_l=p_l=0$ at $t=0$. Due to the nonlinear nature of the model the
norm distribution spreads, while keeping the total norm $S=\sum_l
(q_l^2+p_l^2)/2=1$ constant. The performance of the various integration
schemes is evaluated by their ability to (a) reproduce the
dynamics correctly, which is reflected in the subdiffusive increase of $m_2(t)$,
(b) keep the values of the two integrals $H_D$ and $S$ constant, as
monitored by the evolution of the absolute relative errors of the
energy $E_r(t)=|(H_D(t)-H_D(0))/H_D(0)|$ and of the norm
$S_r(t)=|(S(t)-S(0))/S(0)|$, and (c) reduce the required CPU time
$T_c(t)$ for the performed computations. Results obtained for the SIs $ABC_4$, $ABC_6$ and $SS_4$ are presented in Figure \ref{fig:2}.
A more extended numerical analysis including high-performance non-symplectic schemes can be found in \citep{DNLS}.

All three tested methods are used with different time steps ($\tau_{SS_4}=0.1$, $\tau_{ABC_4}=0.05$ and $\tau_{ABC_6}=0.15$),
chosen in such a way as to provide a comparable error in energy conservation.
In Figure \ref{fig:2} we see that all of these integration methods are able to correctly describe the system's dynamical evolution, since the wave packet's $m_2$ shows practically the same behavior in all cases (Figure \ref{fig:2}a).
Figure \ref{fig:2}b shows, that all integration methods are able to keep the local relative energy error practically constant at $E_r\approx 10^{-5}$. The relative norm error $S_r$, however, grows with time for all applied methods (Figure \ref{fig:2}c) especially for $ABC_4$ and $ABC_6$. The $S_r$ for $SS_4$ increases only slightly in the tested integration interval.
Nevertheless, our results indicate that all methods can keep $S_r$ at acceptable levels (e.g. $S_r\le 10^{-2}$), even for long time integrations such as needed for asymptotic studies of the DNLS model.
The $ABC_4$ and $ABC_6$ SIs require less CPU times compared to the $SS_4$ method (Figure \ref{fig:2}d), with $ABC_6$ showing again the best performance.
From the SIs' performance presented in Figure \ref{fig:2} we conclude that one obtains satisfying results with all symplectic three part split methods.
Regarding the inter-SI competition, we find that using $ABC_6$ with $\tau=0.15$ requires $\sim 1.6$ times less CPU time than the $SS_4$ with $\tau=0.1$.
Our results indicate that the application of efficient triple split SIs allows the accurate  integration of the DNLS over considerably long times. Thus, triple split SIs will possibly allow the community to numerically tackle open questions regarding the asymptotic behavior of wave packets.

\section{\label{sec:conclusion}CONCLUSIONS}
In this work we have collected and tested different symplectic integration algorithms applicable to general Hamiltonians,
which can be split in three integrable parts. After a general introduction to the theory behind symplectic integrators we explicitly explained how one can construct
high order symplectic schemes for such Hamiltonians.
The efficiency of the developed methods was tested on two different Hamiltonian systems,
a simple toy model and the disordered discrete nonlinear Schr\"odinger equation.
We could show that three part split symplectic integrators are very efficient tools for long time integrations of multidimensional Hamiltonian systems
as they tend to preserve conserved quantities and save computational time.

\section*{Acknowledgments}
S.E.~acknowledges the support of the
European Union Seventh Framework Program (FP7/2007-2013) under grant agreement no. 282703. Ch.S.~would like to thank the Max Planck Institute for the
Physics of Complex Systems in Dresden for its hospitality during his
visit in January--February 2013, when part of
this work was carried out. Ch.S.~was supported by the Research
Committee of the Aristotle University of Thessaloniki (Prog.~No
89317), and by the European Union (European Social Fund - ESF) and
Greek national funds through the Operational Program ``Education and
Lifelong Learning'' of the National Strategic Reference Framework
(NSRF) - Research Funding Program: ``THALES. Investing in knowledge
society through the European Social Fund''.

\renewcommand{\refname}{

\end{document}